\documentclass[twocolumn,prb,aps,amsmath,amssymb,floatfix,showpacs]{revtex4}
\usepackage{graphicx}
\usepackage{bm}
\begin{document}
\title{First principles calculations of  X-ray absorption in an
ultrasoft pseudopotentials scheme: from $\alpha$-quartz to high-T$_c$ compounds. }
\author{Christos Gougoussis}
\author{Matteo Calandra}
\author{Ari P. Seitsonen}
\author{Francesco Mauri}
\affiliation{CNRS and Institut de Min\'eralogie et de Physique des Milieux condens\'es, 
case 115, 4 place Jussieu, 75252, Paris cedex 05, France}
\date{\today}

\begin{abstract}
We develop a first-principles scheme based on the continued fraction approach and 
ultrasoft pseudopotentials to calculate K-edge X-ray absorption spectra in solids. 
The method allows for calculations of K-edge X-ray 
absorption spectra in transition metal
and rare-earths compounds with substantially reduced cutoffs respect to the 
norm-conserving case. 
We validate the method by calculating Si and O K-edges in $\alpha$ quartz,
Cu K-edge in Copper and in La$_2$CuO$_4$.
For the case of Si and O edges in $\alpha$ quartz and in Copper we obtain a 
good agreement with experimental data. 
In the Cu K-edge spectra of La$_2$CuO$_4$, a material considered a real challenge for
density functional theory we attribute all the 
near-edge and far-edge peaks to single particle excitations. 
\end{abstract}

\maketitle

\section{Introduction}

With the development of synchrotron radiation sources, 
X-ray absorption spectroscopy (XAS) has become a very
powerful and a widely used technique to investigate structural 
properties and electronic structures in condensed matter 
physics. Since the absorption of X-rays at well-suited energies is 
chemical and orbital selective, it is possible
to probe electronic excitations and explore the local environment 
around the absorbing atom. 
The use of polarized X-rays allows to separate the 
contributions of different atomic orbitals 
through the study of angular dependence of the  spectra.\cite{0953-8984-2-3-018}

K-edge XAS has been used to study the electronic structure of correlated transition metal compounds,  \cite{KotaniBook,shukla:077006,gougoussNiO}, 
to probe the local environment around impurities in crystals \cite{juhin:054105}
and in disordered matter like glass or liquids. For instance, 
XAS plays a crucial role for the understanding of the microscopic structure of water
\cite{wernetscience2004,wang_water_pccp_2006,smith_water_jpcb_2006,prendergast:215502,headgordon_pnas_2006,brancato:107401}. 
The widespread use of XAS as a structural utility and as a probe of the electronic structure requires reliable
theoretical approaches to interpret the measured spectra.

Different theoretical methods are available to calculate XAS. 
The multiplet approach \cite{Cowan1981}, used to calculate pre-edge features for systems with localized final states, 
relies on the solution of a few-sites manybody hamiltonian including several parameters that are suitably chosen
to fit the experimental data. This approach, although providing a full manybody solution to the problem, has three main
shortcomings; (i) it is limited to pre-edge structures and (ii) non-local excitations are hardly taken into account
due to the short-range nature of the considered clusters, (iii) in some 
cases a small variation of the 
hamiltonian parameters leads to substantially different spectra.
The multiple scattering approach\cite{PhysRevA.22.1104,0953-8984-4-43-004,PhysRevB.58.7565} and its extension to 
non muffin-tin potentials have been widely used with success, 
but these methods are not based on first principles and require adjustable parameters
to interpret the experimental data. 
The solution of the Bethe Salpeter equation is the first principle method having
the most satisfactory treatment of many-body effects \cite{ShirleyPRL98}. However
being this method extremely time consuming it only allows for a description of the
pre-edge region of the XAS spectra: the near-edge and far-edge regions cannot
be easily computed.

Density functional theory (DFT) approaches 
 \cite{PhysRevB.66.195107,PhysRevB.50.17953,GiannozziXAS04,Pickard97,Pickard99,GaoPRB08}, 
have been successfully applied to K-edges of weakly correlated materials.  
In a pseudopotential framework, the use of the PAW \cite{PhysRevB.50.17953} method
allows to reconstruct the all electron wavefunction and consequently to
obtain XAS intensities unaffected by the presence of a pseudopotential.
Furthermore the development of a DFT method \cite{PhysRevB.66.195107}
using norm-conserving pseudopotentials
and based on the continued fraction approach permits to obtain XAS spectra up to the
far edge region.
Another advantage of DFT approaches is that they allow for structural
optimization of the local environment around the absorbing atom, a key
issue in the case of impurities or defects.

In time-independent DFT methods, core-hole effects are included in a supercell 
approach by generating a pseudopotential with a core-hole
in the desired atomic core level. 
While this method works very well for weakly correlated system, 
in the presence of moderate or
strong correlation it has two main shortcomings. 
The first is the unsatisfactory treatment of electron-electron
interaction in the DFT functional. The second is the large increase in 
computational time when dealing with transition metals and rare-earths mainly 
related to the huge kinetic energy cutoffs involved and the need 
to simulate large supercells with reduced symmetry .
\footnote{The presence
of a core-hole in the pseudopotential of the absorbing atom 
reduce the symmetry of the crystal. }
A partial remedy for the lack of correlation effects is the use of the 
DFT+U approximation\cite{PhysRevB.44.943}.
Recently the method of ref. \cite{PhysRevB.66.195107} was generalized to the 
DFT+U approximation \cite{gougoussNiO}.
It was shown the DFT+U dramatically improves the agreement with 
experimental data in the pre-edge region 
of correlated 3d transition metal compounds \cite{gougoussNiO}.
Still the second problem holds, namely the huge cutoffs needed to simulate transition 
metal and rare-earth compounds require a substantial computational time. 

In this work we solve this problem developing a method to calculate XAS in an 
Ultrasoft pseudopotential \cite{PhysRevB.41.7892} scheme and relying on 
the continued fraction approach.  
US pseudopotentials allow for small cutoffs (20-40 Ry) even
for transition-metals and rare earth systems, contrary to norm-conserving ones. 
Thus the use of these pseudopotentials
reduces the computational cost of the supercell calculation by an order of magnitude. 
The drawback is that
the continued fraction scheme developed in ref. \cite{PhysRevB.66.195107} does not 
apply if at least one ultrasoft 
pseudopotential (not necessary the absorbing atom) is present in the calculation. 
For this reason we reformulate completely the continued fraction and the corresponding 
lanczos approach in a way that is suitable for Ultrasoft pseudopotentials. 
We then apply the method to Si and O K-edge in $\alpha-$quartz, to Cu K-edge 
in copper and in La$_2$CuO$_4$ and
compare the results with available experimental data.

The structure of the paper is the following. 
In secs. \ref{section:xas} and \ref{section:paw} we remind the general 
expression of the XAS cross section within the projector augmented wave formalism. 
In sec. \ref{section:us} and \ref{section:lanczos} we develop the continued fraction 
approach in the case of ultrasoft pseudopotentials, 
and finally in sec. \ref{section:apps} we apply the method to the aforementioned systems.

\section{X-ray absorption cross-section}
\label{section:xas}

The XAS cross section is \cite{0953-8984-2-3-018} :

\begin{equation}
 \label{eq:exprcrosssec1}
 \sigma(\omega) = 4 \pi^2 \alpha \hbar \omega \sum_{f} \left| M_{i\rightarrow f}\right|^2 \delta(E_f-E_i-\hbar \omega)
\end{equation}

where $\hbar \omega$ is the incident photon energy, $\alpha$ is the fine-structure constant, and $M_{i\rightarrow f}$ is the transition amplitude between the initial state $\left| \psi_i \right\rangle $ of energy $E_i$ and the final state $\left| \psi_f \right\rangle $ of energy $E_f$. 
In a single particle approach, the many body $\left|\psi_i\right\rangle$ and $\left|\psi_f\right\rangle$ are replaced 
by single-particle states. Since we consider K and L$_{1}$ edges, 
$\left|\psi_i\right\rangle$ can be either 
the 1s or the 2s atomic core state in the absence of a core-hole. 
The final state $\left|\psi_f\right\rangle$ in the presence of a core-hole 
is obtained in an all-electron first-principles calculation.

In a single particle approach and in the electric quadrupole approximation, 
the transition amplitude is given by the matrix element
\begin{equation} 
M_{i\rightarrow f} = \left\langle \psi_f \middle| \mathcal{D} \middle| \psi_i \right\rangle
\label{eq:Mifdef}
\end{equation}
with :
%\begin{equation}
%\label{eq:exprmif1}
%\left\lbrace  \begin{array}{c} M_{i\rightarrow f} = \left\langle \psi_f \middle| \mathcal{D} \middle| \psi_i \right\rangle \\
% \mathcal{D}= \mathbf{\hat{\epsilon}}\cdot\mathbf{r}+\frac{\textrm{i}}{2} (\mathbf{\hat{\epsilon}}\cdot\mathbf{r})( \mathbf{k}\cdot\mathbf{r} ) \end{array} \right. 
%\end{equation}
\begin{equation}
\label{eq:exprmif1}
 \mathcal{D}= \mathbf{\hat{\epsilon}}\cdot\mathbf{r}+\frac{\textrm{i}}{2} (\mathbf{\hat{\epsilon}}\cdot\mathbf{r})( \mathbf{k}\cdot\mathbf{r} )
\end{equation}
where $\hat{\epsilon}$ and $\mathbf{k}$ are the popularization vector and the wave vector of the incident beam and ${\bf r}$ is the electron coordinate.

\section{X-ray absorption cross-section in a PAW formalism}
\label{section:paw}

In a first-principles pseudopotential approach, the calculated wavefunction
is $| \tilde{\psi}_f \rangle$, namely the pseudowavefunction of the crystal 
obtained at the end of the self consistent field run. 
In order to get the all-electron wavefunctions 
$\left| \psi_f \right\rangle$ needed in Eqs. \ref{eq:exprcrosssec1},
\ref{eq:exprmif1}, all electron reconstruction needs to be performed.
This is achieved in the framework of the PAW method \cite{PhysRevB.50.17953}. 
In this approach the all electron wavefunctions $\left| \psi \right\rangle$ are related to the pseudo-wavefunctions 
$| \tilde{\psi} \rangle$ though the linear operator $\mathcal{T}$ :
\begin{equation}
 \label{eq:pawdeft}
 \left| \psi \right\rangle=\mathcal{T} | \tilde{\psi} \rangle
\end{equation}

The $\mathcal{T}$ operator is written as a sum of local contributions centered around each atomic site $\mathbf{R}$ : 
\begin{equation}
\mathcal{T}=1+\sum_{\mathbf{R}}\mathcal{T}_\mathbf{R}
\end{equation}
The local operators $\mathcal{T}_\mathbf{R}$ act only within the so-called augmentation 
regions $\Omega_\mathbf{R}$ centered on 
atomic sites. Following ref.  \cite{PhysRevB.50.17953}, we introduce the all electron (pseudo) 
partial waves $\left|\phi_{\mathbf{R},n}\right\rangle$  
($|\tilde\phi_{\mathbf{R},n}\rangle$), and the projector functions $\left\langle \tilde{p}_{\mathbf{R},n}\right|$ 
that satisfy the conditions \cite{cit29cab} :
\begin{eqnarray}
 \tilde\phi_{\mathbf{R},n}(\mathbf{r})&=&\phi_{\mathbf{R},n}(\mathbf{r}) \textrm{ outside }\Omega_\mathbf{R}\\
\left\langle\tilde{p}_{\mathbf{R},n}\middle|\tilde\phi_{\mathbf{R'},n'}\right\rangle&=&\delta_{\mathbf{R} \mathbf{R}'}\delta_{n n'}
\end{eqnarray}
The wavefunctions $\left|\phi_{\mathbf{R},n}\right\rangle$ and $|\tilde\phi_{\mathbf{R},n}\rangle$ respectively form a basis for valence states, which means that any function $\chi_{\mathbf{R}}$ that vanishes outside $\Omega_R$ is expanded as :
\begin{equation}
\label{eq:pawbasis}
 \sum_n \left| \tilde{p}_{\mathbf{R},n} \right\rangle \left\langle \tilde\phi_{\mathbf{R},n} \middle| \chi_{\mathbf{R}} \right\rangle  = \left| \chi_{\mathbf{R}} \right\rangle
\end{equation}
Then the  operator $\mathcal{T}$ is written as :
\begin{equation}
\label{eq:pawexprt}
 \mathcal{T}= \mathbf{1}+\sum_{\mathbf{R},n}\left(\left|\phi_{\mathbf{R},n} \right\rangle -\left|\tilde\phi_{\mathbf{R},n}\right\rangle \right) \left\langle \tilde{p}_{\mathbf{R},n}\right|
\end{equation}
Substituting eq. \ref{eq:pawexprt} in eq. \ref{eq:pawdeft} and eq. \ref{eq:pawdeft} in
eq. \ref{eq:Mifdef}, leads to :
\begin{eqnarray}
M_{i\rightarrow f}&=&\left\langle \tilde\psi_f \middle| \mathcal{D} \middle| \psi_i \right\rangle+\sum_{\mathbf{R},n} \left\langle \tilde\psi_f \middle|\tilde{p}_{\mathbf{R},n}\right\rangle \left\langle  \phi_{\mathbf{R},n}\middle| \mathcal{D} \middle| \psi_i \right\rangle \nonumber \\
&&-\sum_{\mathbf{R},n} \left\langle \tilde\psi_f \middle|\tilde{p}_{\mathbf{R},n}\right\rangle \left\langle  \tilde\phi_{\mathbf{R},n}\middle| \mathcal{D} \middle| \psi_i \right\rangle  \label{eq:paw1}
\end{eqnarray}
Since the initial wavefunction $\psi_i$ is localized on the absorbing atom 
(located at $\mathbf{R}_0$), the terms having ${\bf R}\ne{\bf R_0}$ 
can be neglected in eq. \ref{eq:paw1} to obtain :
\begin{equation}
M_{i\rightarrow f}=\left\langle \tilde\psi_{f} \middle| 
\tilde\phi_{\mathbf{R}_0}\right\rangle
\end{equation}
with :
\begin{equation}
\label{eq:pawphi0}
\left| \tilde\phi_{\mathbf{R}_0}\right\rangle= 
\sum_n \left| \tilde{p}_{\mathbf{r}_0,n} \right\rangle  
\left\langle \phi_{\mathbf{r}_0,n}\middle| \mathcal{D}\middle| \psi_i \right\rangle \,.
\end{equation}
Replacing this matrix element in the XAS cross-section leads to:
\begin{equation}
\label{eq:pawcrosssec2}
\sigma(\omega) = 4 \pi^2 \alpha \hbar \omega \sum_{f} \left|\left\langle \tilde\psi_{f} \middle| \tilde\phi_{\mathbf{R}_0}\right\rangle \right|^2 \delta(E_f-E_i-\hbar \omega)
\end{equation}

Thus Eq. \ref{eq:pawcrosssec2} express the XAS cross-section in terms of single particle states obtained from a 
pseudopotential calculation. Note that in Eq. \ref{eq:pawphi0} there is an infinite number of projectors. 
Practically only a few projectors are needed to achieve convergence.

\section{XAS in an ultrasoft pseudopotential scheme}
\label{section:us}

In an ultrasoft scheme the norm of the pseudo partial waves are different from the norm of the corresponding
all-electron partial waves. For this reason it is customary to define\cite{YatesPRB07} 
the integrated augmentation charges $q_{\bf R,nm}$ as:
\begin{eqnarray}
q_{{\bf R},nm}=\langle \phi_{{\bf R},n}|\phi_{{\bf R},m}\rangle - \langle {\tilde \phi}_{{\bf R},n}|{\tilde \phi}_{{\bf R},m}\rangle
\label{eq:aug_charges}
\end{eqnarray}
The the $S$ operator defined in the ultrasoft scheme \cite{PhysRevB.41.7892} is then:
\begin{eqnarray}
S=\openone + \sum_{{\bf R},m,n}|{\tilde p}_{{\bf R}n} \rangle q_{{\bf R},nm} \langle {\tilde p}_{{\bf R}m}|=\openone+\sum_{\bf R} Q_{\bf R}
\end{eqnarray}

The pseudo hamiltonian $\tilde{H}$ and the pseudo eigenfunctions $|\tilde{\psi_f}\rangle$
 satisfy the following equation\cite{PhysRevB.41.7892} :
\begin{equation}
\label{eqvpus1}
 \tilde{H} \left|\tilde{\psi}_f\right\rangle =E_f S \left|\tilde{\psi}_f\right\rangle
\end{equation}

Multiplication of Eq. \ref{eqvpus1} by $S^{-1/2}$ leads to :
\begin{equation}
\label{eqhspsi2}
S^{-1/2}\tilde{H} S^{-1/2} S^{1/2} \left|\tilde{\psi}_f\right\rangle =E_f S^{1/2} \left|\tilde{\psi}_f\right\rangle
\end{equation}

The following identity holds (for a proof see app. \ref{appendix0}):
\begin{eqnarray}
\pi \sum_{f} \left|\tilde{\psi}_f\right\rangle \delta(E_f-x) 
\left\langle \tilde{\psi}_f \right| 
= \lim_{\gamma\to 0} \Im[{\tilde G}(x)]\label{eqidentitygreen}
\end{eqnarray}
where $x$ is a real number and
\begin{eqnarray}
\tilde G(x) = S^{-1/2} \frac{1}{x-S^{-1/2}\tilde{H}S^{-1/2}-i\gamma} S^{-1/2}\nonumber \\
\end{eqnarray}

Using Eq.\ref{eq:pawcrosssec2} and \ref{eqidentitygreen}, the XAS cross section can finally written in a suitable form for a standard Lanczos procedure :
\begin{equation}
 \sigma(\omega) = 4 \pi \alpha \hbar \omega \lim_{\gamma\to 0}\Im\left[\left\langle\tilde\phi_{\mathbf{R}_0}\middle|{\tilde G}(\omega+E_i)\middle|\tilde\phi_{\mathbf{R}_0}\right\rangle\right]\label{eq:Img}
\end{equation}

where $E_i$ is the energy of the initial state that in a pseudopotential scheme
is undetermined up to an overall constant. In the case of a unit cell having multiple
absorbing sites which are equivalent under the point group symmetry of the crystal,
$E_i$ is the same for all the absorbing atoms and the choice of $E_i$
simply corresponds to a rigid shift of the overall spectrum.
On the contrary, in the case of nonequivalent absorbing sites in
the unit cell, the value of $E_i$ depends on the absorbing site due to the core-level
shift. In this case the choice of $E_i$ is not arbitrary
and a careful determination of the core-level shift is needed \cite{GaoPRB08}.
For simplicity in this work we consider only examples in which there are only
equivalent absorbing sites in the unit cell. The determination of the core-level shift
in the case of multiple nonequivalent absorbing sites will be given elsewhere.  
Thus in what follows we choose the energy
$E_i$ to be the Fermi level, in the metallic case, the highest occupied state,
in the insulating case.

\section{Lanczos procedure}
\label{section:lanczos}

Eq. \ref{eq:Img} can be calculated using the  the Lanczos recursion method 
\cite{Lanczos1952,Lanczos1950,0022-3719-5-20-004,0022-3719-8-16-011}. 
The quantity $\langle \tilde\phi_{\mathbf{R}_0} |\tilde{G}(E_i+\hbar\omega) | \tilde\phi_{\mathbf{R}_0} \rangle$ 
is evaluated using the continued fraction :
\begin{equation}
 \left\langle \tilde\phi_{\mathbf{R}_0}\middle|\tilde{G}(E)\middle|\tilde\phi_{\mathbf{R}_0}\right\rangle=\frac{\left\langle\tilde\phi_{\mathbf{R}_0}\middle|\tilde\phi_{\mathbf{R}_0}\right\rangle}{a_0-E-i\gamma-\frac{b_1^2}{a_1-E-i\gamma-\frac{b_2^2}{\ddots}}}
\end{equation}
where the real numbers $a_i$ and $b_i$ are computed recursively by defining the vectors 
$\left| u_i \right\rangle$ such that:
\begin{eqnarray}
\label{eq:recur1}
\left| u_0 \right\rangle&=&\frac{S^{-1/2}\left|\tilde\phi_{\mathbf{R}_0}\right\rangle}{\sqrt{\left\langle\tilde\phi_{\mathbf{R}_0}\middle| S^{-1} \middle| \tilde\phi_{\mathbf{R}_0}\right\rangle}}\nonumber \\
S^{-1/2}\tilde{H}S^{-1/2} \left| u_i \right\rangle&=&a_i \left| u_i \right\rangle+ b_{i+1} \left| u_{i+1} \right\rangle+b_i \left| u_{i-1} \right\rangle \nonumber 
\end{eqnarray}
The $a_i$ and $b_i$ coefficients are defined as:
\begin{eqnarray}
a_i&=&\left\langle u_i \middle| S^{-1/2}\tilde{H}S^{-1/2} \middle|  u_i\right\rangle\\
b_i&=&\left\langle u_i \middle| S^{-1/2}\tilde{H}S^{-1/2} \middle|  u_{i-1}\right\rangle
\end{eqnarray}
This is essentially a standard lanczos process where the initial vector 
is $|u_0\rangle$ and the Hamiltonian $\tilde{H}$ is replaced by 
$S^{-1/2}\tilde{H}S^{-1/2}$. However this is not the more
efficient way to carry out the lanczos chain since two multiplications 
by $S^{-1/2}$ are involved
and the S matrix is of the same order as the Hamiltonian, 
namely the dimension is given by the number
of plane waves in the calculation (the kinetic energy cutoff). Thus any application of 
$S^{-1/2}$ costs as much as the application of $\tilde{H}$.

A more efficient way to implement the lanczos process is obtained by defining the auxiliary vectors $|t_i\rangle$, namely:

\begin{equation}
 \left| t_i \right \rangle=S^{1/2}\left| u_i \right \rangle
\end{equation}
Using this definition, the lanczos process can now directly carried out on the $|t_i\rangle$ vectors as:

\begin{eqnarray}
\label{eq:recur2}
\left| t_0 \right\rangle&=&\frac{\left|\tilde\phi_{\mathbf{R}_0}\right\rangle}{\sqrt{\left\langle\tilde\phi_{\mathbf{R}_0}\middle| S^{-1} \middle| \tilde\phi_{\mathbf{R}_0}\right\rangle}}\nonumber \\
\tilde{H}S^{-1} \left| t_i \right\rangle&=&a_i \left| t_i \right\rangle+ b_{i+1} \left| t_{i+1} \right\rangle+b_i \left| t_{i-1} \right\rangle\nonumber 
\end{eqnarray}
where the new lanczos vectors $|t_i\rangle$ are no longer orthogonal but $\langle t_i|S^{-1} |t_j\rangle=\delta_{i,j}$.
If during the lanczos chain the vectors $|{\tilde t}_i\rangle = S^{-1}|t_i\rangle$ are stored then
the $a_i$ and $b_i$ coefficients can be defined as:
\begin{eqnarray}
a_i&=&\left\langle {\tilde t}_i \middle|\tilde{H} \middle|  {\tilde t}_i\right\rangle\\
b_i&=&\left\langle {\tilde t}_i \middle|\tilde{H} \middle|  {\tilde t}_{i-1}\right\rangle
\end{eqnarray}
Now, each iteration needs only one multiplication by $S^{-1}$, one multiplication by $\tilde{H}$, 
and four lanczos vectors stored in memory, namely $|t_{i-1}\rangle$, $|{\tilde t}_{i-1}\rangle$,
$|t_{i}\rangle$ and $|{\tilde t}_{i}\rangle$.

To achieve an efficient implementation of the lanczos process, particular care needs 
to be taken in inverting the $S$ matrix to calculate $S^{-1}$. 
Direct inversion of the $S$ matrix is unfeasible being the order of the matrix
given by the number of plane waves. 
Using the definition of $S$ in terms of the $N_p$ ultrasoft projectors, 
the calculation of $S^{-1}$ can be performed 
very efficiently by simple products and inversions of matrices of the order of 
$N_p\times N_P$, as it was demonstrated in refs. \cite{Hasnip06,WalkerJCP07}. 
In order to have a complete description of the method used
we recall the main passages of the demonstration of ref.
\cite{Hasnip06,WalkerJCP07} in the appendix \ref{appendixa}.

\section{Applications}
\label{section:apps}

The developed method is now applied to silicon and oxygen K-edges in $\alpha$-quartz, 
to Cu K-edge in copper and in La$_2$CuO$_4$. 
Density functional theory calculations are performed using the Quantum-Espresso
package \cite{QE} and the Generalized Gradient Approximation \cite{PhysRevLett.77.3865}.
In the case of La$_2$CuO$_4$ we use the Spin Polarized Generalized Gradient Approximation.
The developed continued-fraction approach to deal with US pseudopotentials is implemented
in the XSpectra package \cite{XSpectra} and distributed with 
the current CVS version of the Quantum-Espresso code.
Occupied states are eliminated from the spectrum using the method 
of ref. \cite{PhysRevB.54.7334}. The zero of energy is determined from the
self-consistent calculation on a supercell in the presence of a core-hole.
In the metallic case we chose the Fermi level while in the insulating case
the highest occupied state.
It is important to notice that an insulator can become metallic in the supercell calculation 
due to core-hole attraction. In this case the elimination of the occupied states is somewhat ill
defined, as it is in metallic systems. When this occurs, the pre-edge features can be incorrect.
This is the case in La$_2$CuO$_4$.
Further technical details of the calculations are given in each
subsection.

\subsection{SiO$_2$ ($\alpha$-quartz)}

SiO$_2$ ($\alpha$-quartz) is a dichroic compound with a hexagonal unit cell and lattice parameters $a=4.9141 \AA$ and $c=5.4060 \AA$ \cite{lager:6751}. 
The dipolar cross section $\sigma$ has the following angular dependence 
\cite{0953-8984-2-3-018} :
\begin{equation}
\label{eq1}
\sigma(\mathbf{\epsilon})=\cos^2(\theta)\sigma_{\parallel}+\sin^2(\theta)\sigma_{\perp}
\end{equation}
where $\mathbf{\epsilon}$ is the polarization vector and $\theta$ is the angle between the c axis and $\mathbf{\epsilon}$. 

The charge density calculation was performed with a $2\times2\times2$ supercell containing $72$ atoms. Electronic 
integration was performed using only the ${\bf \Gamma}$ point. 
We used a 20 Ry kinetic energy cutoff  and a 150 Ry cutoff for the charge density, 
to be compared to the 70 Ry kinetic energy cutoff  needed in a standard norm conserving 
pseudopotentials calculation \cite{PhysRevB.66.195107}. 
The electronic integration in the continued fraction  calculation using the lanczos 
method was performed using a centered $3\times3\times3$ \textbf{k}-points grid of the
$72$ atoms supercell. 
Two projectors per channel were used in the PAW reconstruction. 
The core-hole width was taken constant and set to $0.8$ eV for Si K-edge and 1 eV for O K-edge. 
The continued fraction calculation needed around 400 iterations per \textbf{k}-point. 
We performed a calculation with norm-conserving pseudopotentials in 
which around 600 iterations were 
needed for comparable accuracy, like in previous work\cite{0953-8984-2-3-018}. 
Thus the number of iterations needed is smaller, 
due to the smaller cutoff energy.

\begin{figure}[!h]
\includegraphics[width=\columnwidth]{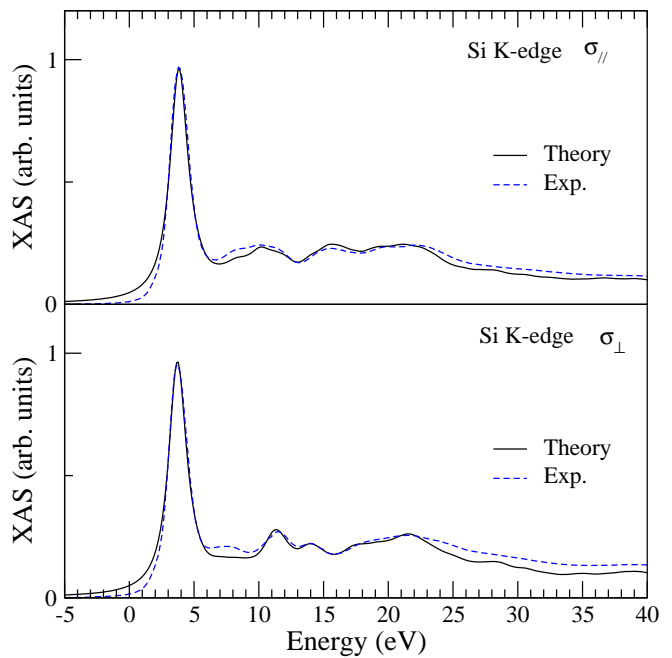}
\caption{ Experimental \cite{PhysRevB.66.195107} and calculated Si K-edge in 
$\alpha$-quartz.  $\sigma_\|$ is the polarization along the c axis, 
while $\sigma_\perp$ is the in-plane polarization.}
\label{fig:SiO2_SiK}
\end{figure}

\begin{figure}[!h]
\includegraphics[width=\columnwidth]{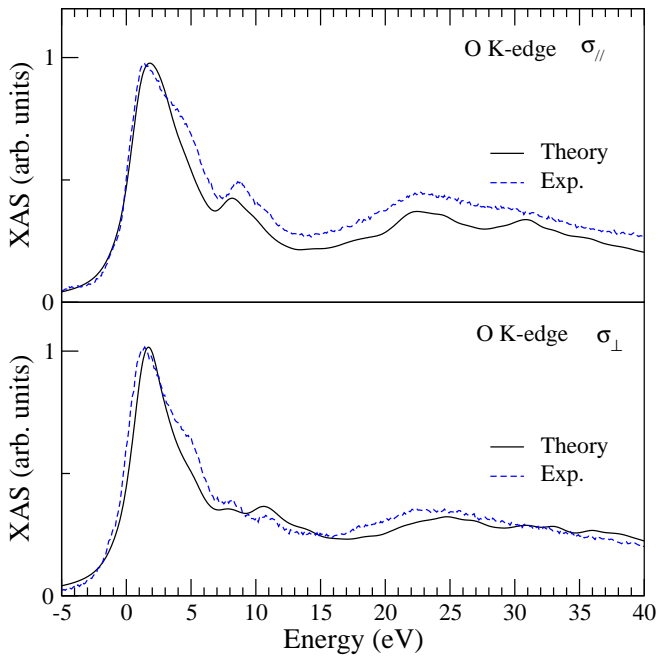}
\caption{ Experimental  \cite{PhysRevB.66.195107} and calculated O K-edge 
in $\alpha$-quartz. $\sigma_\|$ is the polarization along the c axis, 
while $\sigma_\perp$ is the in-plane polarization.}
\label{fig:SiO2_OK}
\end{figure}

Our results are presented in figures \ref{fig:SiO2_SiK} and \ref{fig:SiO2_OK}. These results are in perfect
agreement with those presented in ref. \cite{PhysRevB.66.195107} and obtained with norm-conserving pseudopotentials. 
The experimental Si k-edge XAS cross section is very well reproduced for both polarizations despite a too weak peak 
around 7 eV. A good agreement between theory and experiment for O K-edge is obtained.

The comparison between the ultrasoft pseudopotential and the norm conserving pseudopotential calculations on 
$\alpha$-quartz validates our implementation of XAS using ultrasoft pseudopotentials. 

\subsection{Copper}

Pure copper at room temperature crystallizes in the fcc structure with lattice parameter $3.601 \AA$ \cite{otte:1536}. 
The copper K-edge XAS cross section was calculated on a converged $3\times3\times3$ 
supercell containing 27 atoms. Electronic integration was performed over
a $10\times10\times10$ uniform k-point grid for both charge density and XAS calculations, with a 30 Ry kinetic energy cutoff  
and a 500 Ry charge-density cutoff.
 
\begin{figure}[!h]
\includegraphics[width=\columnwidth]{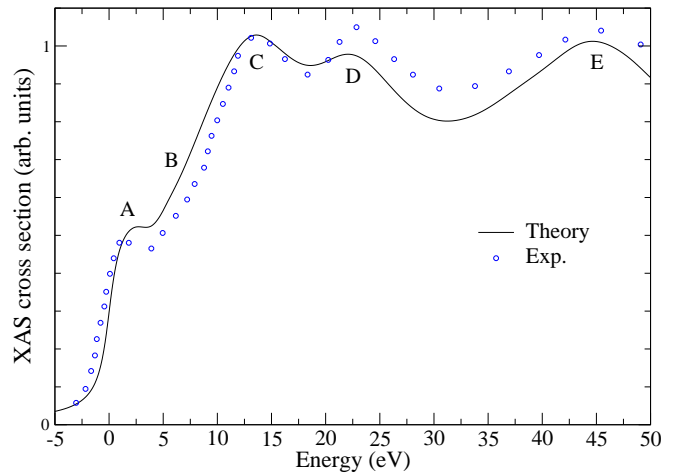}
\caption{Calculated Cu K-edge in copper compared 
to experimental
data from ref. \cite{PhysRevA.64.062506}. 
The Lorentzian $\gamma$ broadening parameters is variable and runs 
from 1 to 4 eV.  }
\label{fig:cu1}
\end{figure}
The calculated XAS cross section is in good agreement with experimental data 
(fig. \ref{fig:cu1}). The features A, B, C, D and E are correctly reproduced, however
 the peak A (shoulder) is shifted of about 1 eV to higher energies. 
This shift is also present in finite difference method calculations 
\cite{PhysRevB.63.125120}. 
Since copper is metallic, the description of the A peak 
is difficult, because the core-hole attraction can drag some states below the Fermi 
energy.

\subsection{La$_2$CuO$_4$}

La$_2$CuO$_4$ is the parent compound of high Tc superconductors. 
It is an antiferromagnetic correlated insulator 
considered a challenge for density functional theory. 
Furthermore it requires large cutoff energies to be simulated with
norm-conserving pseudopotentials. Thus it is an ideal test for
our approach.

At low temperatures La$_2$CuO$_4$ present an weak orthorhombic 
distortion of the tetragonal structure. 
In our calculation we neglect the orthorhombic distortion and 
consider the tetragonal structure
having $a=5.357$ \AA and $c=13.143$ \AA \cite{reehuis:144513}.
Under this assumption, the Cu K-edge XAS dipolar cross section can be described as a linear combination of the 
cross-section having in-plane polarization ($\sigma_{\perp}$) and of that having polarization along the
c-axis ($\sigma_{\parallel}$),as expressed in eq. \ref{eq1}.

We treat correlation effects the framework of the the spin polarized GGA+U 
\cite{PhysRevLett.77.3865,cococcioni:035105} approximation,
where U is the Hubbard parameter on Cu 3d states. The U parameter is $9.6$ eV, as 
calculated from first principles using a linear response scheme 
\cite{cococcioni:035105,kulik:103001}. 
We use ultrasoft pseudopotentials for all 
atomic species leading to a $30$ Ry kinetic energy cutoff and a $200$ Ry charge density 
cutoff. The kinetic energy cutoff used with ultrasoft pseudopotentials
has to be compared with the more than $150$ Ry kinetic energy
cutoff needed in the case of norm conserving pseudopotentials. 
We use two PAW projectors per channel and non-linear core correction in the Cu
pseudopotential. By calculating XAS on the antiferromagnetic unit cell
we have checked that 
the inclusion of semicore states does not affect the result. 
We used a $1\times1\times1$ supercell of the antiferromagnetic crystal cell 
containing 14 atoms.
For the electronic integration we use a uniform $6\times6\times6$  k-point mesh both 
for the charge density and for the continued fraction calculation.
We have verified that the result is unaffected by the use of larger supercells.

Experimentally, La$_2$CuO$_4$ is an insulator with a gap around 2 eV 
\cite{PhysRevB.37.7506}, and exhibits an antiferromagnetic order 
\cite{PhysRevLett.58.2802} with a magnetic momentum on copper atoms around $0.5 \mu_B$. 
Our CGA+U electronic structure calculation gives a 0.5 eV charge-transfer gap 
and a magnetization of $0.58\mu_B$. 
\begin{figure}[!h]
\includegraphics[width=0.9\columnwidth]{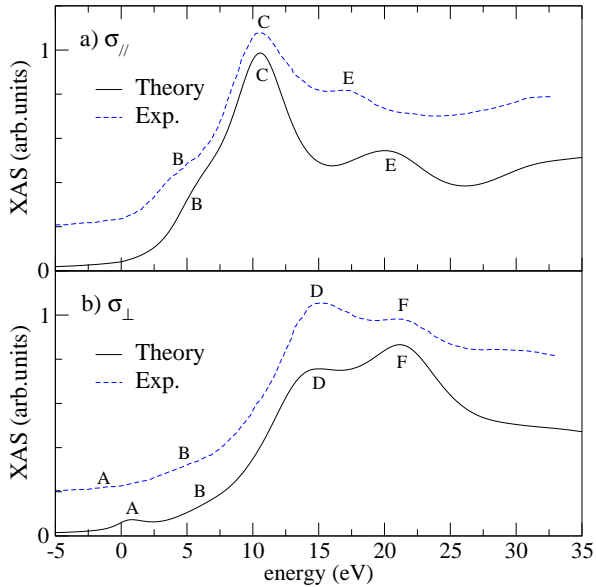}
\caption{Experimental\cite{PhysRevB.45.8091} and calculated Cu K-edge in La$_2$CuO$_4$.
The calculated cross section was obtained with a U=9.6 eV parameter on Cu 3d states.
The Lorentzian $\gamma$ broadening parameters 
varies linearly between 1 eV in the pre-edge region and 4 eV in the far-edge
region. 
$\sigma_\|$ indicates the polarization along the c axis while $\sigma_\perp$ the in-plane
 one.}
\label{fig:lacuo1}
\end{figure}

The results of the CGA+U Cu K-edge XAS calculations are presented in fig. \ref{fig:lacuo1}. 
When the polarization is parallel to the CuO$_2$ planes the energy position of the 
different peaks is well reproduced by our calculation. In particular B, D,and F are at
the correct energy position, however the intensity of peak D is underestimated.
In the pre-edge region the peak A is shifted to higher energy. 
The A peak is not well described in our calculation since, due to the underestimation
of the electronic gap and overestimation of core-hole attraction, 
the system becomes metallic when a core-hole is included in the calculation.
In particular, while the system in the absence of a core-hole is insulating, when we add
the core-hole in the supercell calculation, we obtain a metallic system. As a consequence it becomes
impossible to distinguish between occupied and empty states.
The intensity of peak A is crucially affected. 
The system is formally in a $|3d^{10}\underline{L}\rangle$ state with no empty d-states and consequently
no quadrupolar pre-edge. On the contrary it is known that a weak quadrupolar pre-edge is present in experiments
\cite{shukla:077006}.

The nature of peak B has been widely discussed. 
It has been alternatively assigned to shakedown  
$1s^23d^9L \longrightarrow 1s^13d^{10} \underline{L}$ processes \cite{PhysRevB.22.2767}, 
and to empty p states of the absorbing atom \cite{PhysRevB.45.8091,PhysRevB.41.131}. 
Here we unambiguously attribute the peak B  to  empty p states and thus 
this excitation is single particle in nature. 

\begin{figure}[]
\includegraphics[width=0.9\columnwidth]{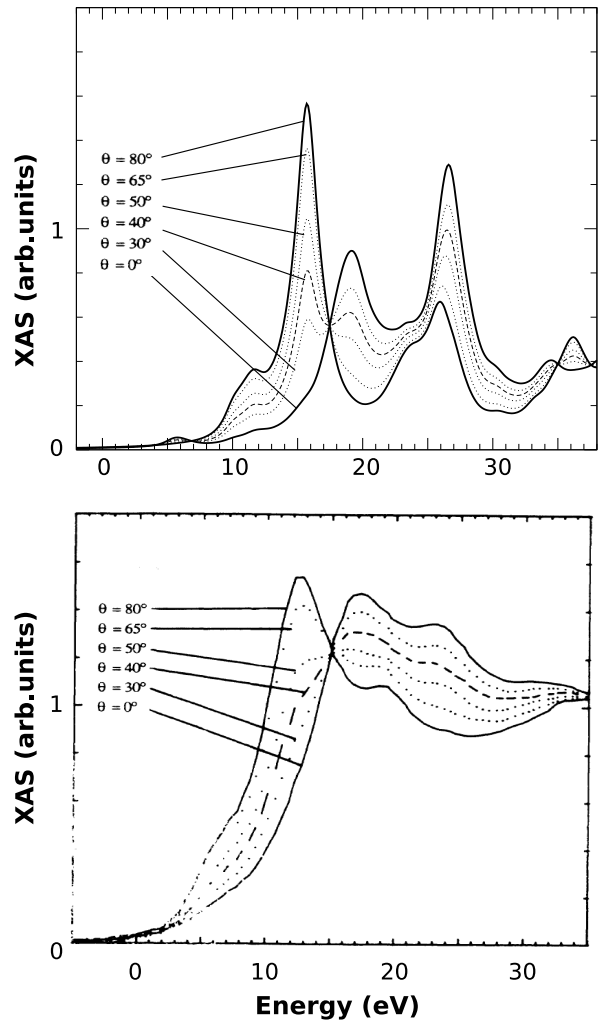}
\caption{Angular dependence of La$_2$CuO$_4$ Cu K-edge XAS dipolar cross section compared with experimental
data \cite{PhysRevB.45.8091}. The angle between the CuO$_2$ layer and the c-axis is labeled $\theta$, so that 
$\theta=90^{o}$ corresponds to $\epsilon$  along the c-axis. 
The Lorentzian $\gamma$ broadening parameters is  0.8 eV. The core-hole
width has been artificially reduced in the calculation to show the presence of different peaks.}
\label{fig:lacuo4angdep}
\end{figure}

In the spectra having polarization along the c-axis the intensities are in
better agreement with experimental data. However the 
peak E is substantially shifted to higher energies (3 eV).
More insight on this issue can be obtained by considering the angular dependence 
of the spectra and by 
comparing it to available experimental data \cite{PhysRevB.45.8091}, 
as shown in Fig.\ref{fig:lacuo4angdep}.

In the theoretical calculation of Fig. \ref{fig:lacuo4angdep} we have substantially reduced the peak
linewidth to 0.8 eV to emphasize the multi structured form of the different peaks.
In particular it is seen that the peak at 20 eV  (labeled peak E in Fig. \ref{fig:lacuo1}) is actually composed
of two different peaks. The high energy one is at the correct energy position while the low energy one is shifted
to higher energy with respect to the experiment. The incorrect position of the lowest energy peak substantially affects the intensity
in the 10-20 eV region for both polarizations 
and is responsible for the disagreement with the experimental data.
Indeed if this peak was at low energy the intensity of peak D in Fig. \ref{fig:lacuo1}) would also increase and a better
agreement with experiment would be obtained for both polarizations.

Kosugi et al. \cite{PhysRevB.41.131} attributed the decoupling of pre-edge and near-edge
features present in K-edge XAS on powder samples to charge-transfer multi-determinant effects. In particular
it was shown that the main edge and the B peak are decoupled in the isotropic spectra. However
the pre-edge and edge structures occur at different energies when the polarization is in the plane
or out of plane. We find  that the decoupling indeed originates from the average over different
polarization of the single-particle spectra. Thus the doubling of edge and pre-edge peaks 
are not due to multi-determinant charge transfer effects. 

Tolentino et al. \cite{PhysRevB.45.8091} assigned peaks C and E 
to $3d^{10} \underline{L}$ transitions, and peaks D and F to $3d^9L$ transitions,
implying that the XAS spectrum includes multi-determinant effects. If this would be
the case, being our calculation single determinant in nature, a single peak
should be seen in both directions. This is however not the case and our calculation
correctly reproduces the main experimental features except for an energy shift for
the peak E. Consequently the near-edge and far-edge structures detected in K-edge XAS
of La$_2$CuO$_4$ are all single particle in origin.
Concerning the E peak, it corresponds to a single particle excitation that
is 3 eV shifted with respect to experiments. This shift can be due to an incorrect description
of the hybridization between Cu 4p states of the absorbing atom and La states.
La states are indeed hard to describe in a single particle approach for their intrinsic
correlated nature. 
We believe that, despite the 3 eV shift of the E-peak, the single particle origin
of the C,D,E,F peaks is definitely clarified.
To have better insight on the subject it would be interesting to study the case of
Ca$_{2-x}$CuO$_2$Cl$_2$ \cite{YamadaPRB05} 
since no rare-earths are present in the system and consequently the E peak should be at
the right position.

\section{Conclusion}

A DFT-based continued fraction method using ultrasoft pseudopotentials to calculate the X-ray absorption spectra is presented. 
Our implementation, relying on ultrasoft pseudopotentials, 
is an order of magnitude faster than
preceding implementations based on norm-conserving one 
\cite{PhysRevB.66.195107,gougoussNiO}. Indeed the bulk of the calculation
is the determination of the self-consistent charge density for a supercell in
the presence of a core-hole. Since Ultrasoft pseudopotentials allow for substantially
smaller cutoffs, the computation cost is strongly reduced.
Furthermore the number of iteration in the continued fraction is reduced
and convergence is then faster than the norm-conserving case.

We validate the method by calculating silicon and oxygen K-edges of alpha-quartz,  
Cu k-edge in bulk metallic copper
and Cu K-edge in La$_2$CuO$_4$. 
In the case of weak to intermediate correlation 
(silicon and oxygen K-edges of alpha-quartz and Cu K-edge in metallic copper) 
we obtain a good agreement with experimental data. 
The description of XAS spectra of strongly correlated compounds as
La$_2$CuO$_4$ is typically considered a challenge for DFT-based method.
Nevertheless we were able to attribute all the  single particle peaks
(B, C, D, F in Fig. \ref{fig:lacuo1}). We then solve the
long-standing \cite{PhysRevB.22.2767,PhysRevB.41.131,PhysRevB.45.8091} 
discussion on the attribution of the near-edge
and far-edge features in La$_2$CuO$_4$.

\section{Acknowledgements}
We acknowledge fruitful discussion with A. M. Saitta, R. Gebauer, 
D. Cabaret, Ch. Brouder, Ph. Sainctavit, N. Marzari
and D. Ceresoli. 
Calculations were performed at the IDRIS supercomputing center (project 081202). 

\appendix

\section{Proof of Eq. \ref{eqidentitygreen}}

\label{appendix0}

We proof that for $x$ real number the following holds:
\begin{eqnarray}
& &\pi \sum_{f} \left|\tilde{\psi}_f\right\rangle \delta(E_f-x) \left\langle \tilde{\psi}_f \right| =\nonumber\\
&=& \lim_{\gamma\to 0}\Im\left[S^{-1/2} \frac{1}{x-S^{-1/2}\tilde{H}S^{-1/2}-i\gamma} 
S^{-1/2} \right]\,. \nonumber \\
 \label{eqidentitygreen2}
\end{eqnarray}
Since
\begin{eqnarray}
& &\pi \sum_{f} \left|\tilde{\psi}_f\right\rangle \delta(E_f-x) \left\langle \tilde{\psi}_f \right| =\nonumber\\
&=& \lim_{\gamma\to 0}\Im\left[\sum_{f} \left|\tilde{\psi}_f\right\rangle \frac{1}{x-E_f-i\gamma} \left\langle \tilde{\psi}_f \right|\right]\nonumber \\
\end{eqnarray}
we have
\begin{eqnarray}
&&\sum_{f} \left|\tilde{\psi}_f\right\rangle \frac{1}{x-E_f-i\gamma} \left\langle \tilde{\psi}_f \right| =\nonumber\\
&=&\sum_{f} S^{-1/2}\frac{1}{x-E_f-i\gamma}S^{1/2}  \left|\tilde{\psi}_f\right\rangle \left\langle \tilde{\psi}_f \right| \nonumber\\
&=&\sum_{f} S^{-1/2} \frac{1}{x-S^{-1/2}\tilde{H}S^{-1/2}-i\gamma} S^{1/2}  \left|\tilde{\psi}_f\right\rangle \left\langle \tilde{\psi}_f \right| \nonumber\\
&=& S^{-\frac{1}{2}} \frac{1}{x-S^{-\frac{1}{2}}\tilde{H}S^{-\frac{1}{2}}-i\gamma} S^{-\frac{1}{2}} \sum_{f}  S  \left|\tilde{\psi}_f\right\rangle \left\langle \tilde{\psi}_f \right| \nonumber\\
&=& S^{-1/2} \frac{1}{x-S^{-1/2}\tilde{H}S^{-1/2}-i\gamma} S^{-1/2} \label{eq:result} \nonumber
\end{eqnarray}
where in the last equality we used the following property \cite{PhysRevB.41.7892}:
\begin{equation}
 \label{eq:eqid}
 \openone=\sum_{f} \left| \tilde\psi_f \right\rangle \left\langle \tilde\psi_f \right| S=\sum_{f} S \left| \tilde\psi_f \right\rangle \left\langle \tilde\psi_f \right| .
\end{equation}
Eq. \ref{eqidentitygreen2} follows from eq. \ref{eq:result}.

\section{Calculating the $S^{-1}$ matrix}
\label{appendixa}

Following ref. \cite{PhysRevB.41.7892}, the S matrix can be written as :
\begin{equation}
S=\openone+\sum_{i,j}q_{ij}\left|{\tilde p}_i\right\rangle\left\langle {\tilde p}_j\right|
\end{equation}
where $i$ and $j$ are cumulative indexes for ${\bf R}m$ and ${\bf R}n$, respectively.
We assume that $S^{-1}$ can be written as:
\begin{equation}
S^{-1}=\openone+\sum_{i,j}a_{ij}\left|{\tilde p}_i\right\rangle\left\langle {\tilde p}_j\right|
\end{equation}
The $S^{-1}$ matrix satisfies the equation $SS^{-1}=\openone$:
\begin{eqnarray}
SS^{-1}&=&(\openone+\sum_{i,j}q_{ij}\left|{\tilde p}_i\right\rangle\left\langle {\tilde p}_j\right|)
(\openone+\sum_{l,m}a_{lm}\left|{\tilde p}_l\right\rangle\left\langle {\tilde p}_m\right|)\nonumber\\
&=&\openone+\sum_{i,j}\left|{\tilde p}_i\right\rangle\left\langle {\tilde p}_j\right| (q_{ij}+a_{ij}+\sum_{lm}q_{ij}P_{jl}a_{lm})\nonumber\\
&&
\end{eqnarray}
where $P_{jl}=\langle {\tilde p}_j | {\tilde p}_l \rangle$. In matrix form the equation is:
\begin{equation}
 q+a+qPa=0
\end{equation}
whose solution is $a=-(1+qP)^{-1}Q$. Thus $S^{-1}$ can be calculated by inverting matrices of the size of $N_P \times N_P$, 
where $N_P$ is the number of ultrasoft projectors. 
A similar procedure was used in ref.\cite{Hasnip06,WalkerJCP07}.


\begin{thebibliography}{99}
\bibitem{0953-8984-2-3-018} C. Brouder, J. Phys. Cond. Mat. {\bf 2},  701-738 (1990)

\bibitem{KotaniBook} F. de Groot and A. Kotani, {\it Core Level Spectroscopy of Solids}, Taylor and Francis, 2008


\bibitem{shukla:077006} A. Shukla, M. Calandra, M. Taguchi, A. Kotani, G. Vanko and S.-W. Cheong, Phys. Rev. Lett. {\bf 96}, 077006 (2006)

\bibitem{gougoussNiO} C. Gougoussis, M. Calandra, A. Seitsonen, Ch. Brouder, A. Shukla, and F. Mauri, Phys. Rev. B {\bf 79}, 045118 (2009)

\bibitem{juhin:054105} A. Juhin, G. Calas, D. Cabaret, L. Galoisy and J.-L. Hazemann,  Phys. Rev. B {\bf 76}, 054105 (2007)

\bibitem{wernetscience2004} P. Wernet, D. Nordlund, U. Bergmann, M. Cavalleri, M. Odelius, H. Ogasawara, L. A.  Naslund, T. K. Hirsch, L. Ojamae, P. Glatzel, L. G. M. Pettersson, A. Nilsson, Science {\bf 304}, 995-999 (2004)

\bibitem{wang_water_pccp_2006} R. L. C. Wang, H. J. Kreuzer and M. Grunze, Phys. Chem. Chem. Phys. {\bf 8},  4744 - 4751 (2006)

\bibitem{smith_water_jpcb_2006} J. D. Smith, C. D. Cappa, B. M. Messer, W. S. Drisdell, R. C. Cohen and  R. J. Saykall, J. Phys. Chem. B {\bf 110}, 20038Ã¢ÂÂ20045 (2006)

\bibitem{prendergast:215502} D. Prendergast and G. Galli,  Phys. Rev. Lett. {\bf 96}, 215502 (2006)

\bibitem{headgordon_pnas_2006} T. Head-Gordon and M. E. Johnson, PNAS {\bf 103}, 7973-7977 (2006)

\bibitem{brancato:107401} G. Brancato, N. Rega and V. Barone, Phys. Rev. Lett. {\bf 100}, 107401 (2008)

\bibitem{Cowan1981} R.D. Cowan, The theory of atomic structure and spectra, University of California Press (1981)

\bibitem{PhysRevA.22.1104} C. R. Natoli,D. K.  Misemer, S. Doniach and F. W. Kutzler, Phys. Rev. A {\bf 22}, 1104--1108 (1980)

\bibitem{0953-8984-4-43-004} L. Fonda,  J. Phys. Cond. Mat. {\bf 4}, 8269-8302 (1992)

\bibitem{PhysRevB.58.7565} L. A. Ankudinov, B. Ravel, J. J. Rehr and S. D. Conradson, Phys. Rev. B {\bf 58}, 7565--7576 (1998)

\bibitem{ShirleyPRL98} E. L. Shirley, Phys. Rev. Lett. {\bf 80}, 794 (1998)

\bibitem{PhysRevB.66.195107} M. Taillefumier, D. Cabaret, A.-M. Flank and F. Mauri, Phys. Rev. B {\bf 66}, 195107 (2002)

\bibitem{PhysRevB.50.17953} P. E. Bl\"ochl, Phys. Rev. B {\bf 50}, 17 953 (1994)

\bibitem{GiannozziXAS04}  Het\'enyi B., De Angelis F., Giannozzi P., and Car R.,  
J. Chem. Phys. {\bf 120}, 8632 (2004)

\bibitem{Pickard97} C. J. Pickard and M. C. Payne, Electron Microscopy and Analysis {\bf 153},  179 (1997) 

\bibitem{Pickard99} P. Rez, J. R. Alvarez, and C. J. Pickard, Ultramicroscopy, {bf 78} 175 (1999)


\bibitem{GaoPRB08} Gao S. P., Pickard C. J., Payne M. C.,  Zhu J., Yuan J., Phys. Rev. B {\bf 77}, 115122 (2008)

\bibitem{PhysRevB.44.943} V. I. Anisimov, J. Zaanen and O. Andersen, Phys. Rev. B {\bf 44}, 943--954 (1991)

\bibitem{PhysRevB.41.7892} D. Vanderbilt, Phys. Rev. B {\bf 41}, 7892--7895 (1990)

\bibitem{cit29cab} The solutions of the radial Schr\"odinger equations for the isolated atom are a natural choice for the all electron partial waves.

\bibitem{YatesPRB07} J. R. Yates, C. J. Pickard, F. Mauri, Phys. Rev. B {\bf 76}, 024401 (2007)

\bibitem{Lanczos1952} C. Lanczos, J. Res. Natl. Bur. Stand. {\bf 49}, 33 (1952)

\bibitem{Lanczos1950} C. Lanczos, J. Res. Natl. Bur. Stand. {\bf 45}, 255 (1950)

\bibitem{0022-3719-5-20-004} R. Haydock, V. Heine and M. J. Kelly, J. Phys. C: Solid State Physics {\bf 5}, 2845-2858 (1972)

\bibitem{0022-3719-8-16-011} R. Haydock, V. Heine and M. J. Kelly, J. Phys. C: Solid State Physics {\bf 8}, 2591-2605 (1975)

\bibitem{Hasnip06} P. J. Hasnip and C. J. Pickard, Computer Physics Communications, {\bf 174}, 24 (2006)

\bibitem{WalkerJCP07} B. Walker and R. Gebauer, J. Chem. Phys. {\bf 127}, 164106 (2007)

\bibitem{QE} P. Giannozzi et al., http://www.quantum-espresso.org

\bibitem{PhysRevLett.77.3865} J.P.Perdew, K.Burke, M.Ernzerhof, Phys. Rev. Lett. {\bf 77}, 3865 (1996)


\bibitem{XSpectra} The XSpectra package by C. Gougoussis, M. Calandra, 
A. Seitsonen and F. Mauri is available under the gnu license in the current CVS version
of the Quantum Espresso code.


\bibitem{PhysRevB.54.7334} Ch. Brouder, M. Alouani and K. H. Bennemann, Phys. Rev. B {\bf 54}, 7334--7349 (1996)

\bibitem{lager:6751} G. A. Lager, J. D. Jorgensen and F. J. Rotella, Journal of Applied Physics {\bf 53}, 6751-6756 (1982)

\bibitem{otte:1536} H.M. Otte, Journal of Applied Physics {\bf 32}, 1536-1546 (1961)

\bibitem{PhysRevA.64.062506} C. T. Chantler, C. Q. Tran, Z. Barnea, D. Paterson, D. J. Cookson and D. X. Balaic,         Phys. Rev. A {\bf 64}, 062506 (2001)

\bibitem{PhysRevB.63.125120} Y. Joly, Phys. Rev. B {\bf 63}, 125120 (2001)

\bibitem{reehuis:144513} M. Reehuis, C. Ulrich, K. Prokes, A. Gozar, G. Blumberg, Seiki Komiya, Yoichi Ando, P. Pattison and B. Keimer, Phys. Rev. B {\bf 73}, 144513 (2006)

\bibitem{cococcioni:035105} M. Cococcioni and S. de Gironcoli, Phys. Rev. B {\bf 71}, 035105 (2005)

\bibitem{kulik:103001} H.J. Kulik, M. Cococcioni, D. A. Scherlis and N. Marzari, Phys. Rev. Lett. {\bf 97}, 103001 (2006)

\bibitem{PhysRevB.37.7506} J. M. Ginder, M. G.  Roe, Y. Song, R. P.  McCall, J. R.  Gaines, E. Ehrenfreund, and  A. J. Epstein,				Phys. Rev. B {\bf 37}, 7506--7509 (1988)

\bibitem{PhysRevLett.58.2802} D. Vaknin, S. K.  Sinha, D. E.  Moncton, D. C.  Johnston, J. M.  Newsam, C. R.  Safinya and H. E. King,	Phys. Rev. Lett.{\bf 58}, 2802--2805 (1987)

\bibitem{PhysRevB.45.8091} H. Tolentino, M. Medarde, A. Fontaine, F. Baudelet, E. Dartyge, D. Guay and G. Tourillon, Phys. Rev. B {\bf 45}, 8091 (1992)

\bibitem{PhysRevB.22.2767} R. Bair and  W. Goddard, Phys. Rev. B {\bf 22} 2767--2767 (1980)

\bibitem{PhysRevB.41.131} N. Kosugi, Y. Tokura, H. Takagi and S. Uchida, Phys. Rev. B {\bf 41}, 131--137 (1990)

\bibitem{YamadaPRB05} I. Yamada, A. A. Belik, M. Azuma, S. Harjo, T. Kamiyama,
Y. Shimakawa, and M. Takano, Phys. Rev. B {\bf 72}, 224503 (2005)






\end{thebibliography}
\end{document}